\documentclass[epj,referee]{svjour}
\usepackage{graphicx}
\usepackage{xspace}

\usepackage{amsmath,amssymb}

\begin{document}

\title{Optimal trapping wavelengths of Cs$_2$ molecules in an optical lattice.}
\author{R.~Vexiau$^{1}$ \and N. Bouloufa$^{1}$ \and M. Aymar$^{1}$ \and J.~G.~Danzl$^2$ \and M. J. Mark$^2$ \and H.-C.~N\"agerl$^2$ \and O. Dulieu$^{1}$}

\institute{$^{1}$Laboratoire Aim\'e Cotton, CNRS, B\^at. 505, Univ Paris-Sud, 91405 Orsay Cedex, France \\
$^{2}$Institut f\"ur Experimentalphysik und Zentrum f\"ur Quantenphysik, Universit\"at Innsbruck
Technikerstra{\ss}e 25, 6020 Innsbruck, Austria }
\date{\today}

\abstract{
The present paper aims at finding optimal parameters for trapping of Cs$_2$ molecules in optical lattices, with the perspective of creating a quantum degenerate gas of ground-state molecules. We have calculated dynamic polarizabilities of Cs$_2$ molecules subject to an oscillating electric field, using accurate potential curves and electronic transition dipole moments. We show that for some particular wavelengths of the optical lattice, called "magic wavelengths", the polarizability of the ground-state molecules is equal to the one of a Feshbach molecule. As the creation of the sample of ground-state molecules relies on an adiabatic population transfer from weakly-bound molecules created on a Feshbach resonance, such a coincidence ensures that both the initial and final states are favorably trapped by the lattice light, allowing optimized transfer in agreement with the experimental observation.
}

%
\maketitle

\section{Introduction}
\label{sec:intro}

The last few years have seen spectacular advances in the field of atomic quantum gases. More recently, it has become a central goal to achieve similar control over each quantum degree of freedom for molecular species This would allow a series of novel fundamental studies in physics and chemistry \cite{carr2009,krems2008}. In particular, for the proposed molecular quantum gas experiments, the molecular ensembles must be prepared at high particle densities combined with ultralow temperatures and the internal degrees of freedom must be controlled at the level of single quantum states. An optical lattice affords exquisite control over the motional wave function of the molecules and lattice-based molecular systems are an ideal starting point for quantum gas studies \cite{jaksch2002,micheli2006,micheli2007} or quantum computation and simulation schemes \cite{demille2002,yelin2006,charron2007} based on ultracold molecules. With each molecule trapped at an individual lattice site, the molecules are shielded from collisional loss during state preparation and manipulation.

The work we report on here is motivated by experiments of the Innsbruck group, which are aimed at controlling molecules in all their degrees of freedom to obtain Bose-Einstein condensates (BEC) of ground-state Cs$_2$ \cite{danzl2008,danzl2010} and ground-state RbCs molecules \cite{lercher2011}. A major step towards this objective has recently been taken when it became possible to produce high-density samples of rovibronic ground-state Cs$_2$ molecules \cite{danzl2010}. A crucial ingredient for these experiments is the presence of an optical lattice. A 3D optical lattice, in its simplest form, is a set of three mutually orthogonal standing wave laser fields. The electric field of these lasers interacts with the atoms or the molecules congregating in the potential minima, which for red-detuned light correspond to the maxima of the standing wave. In order to prepare near quantum degenerate molecular ensembles and in particular high-density ultracold samples of molecules in optical lattices, atoms are first cooled to quantum degeneracy. The BEC of Cs atoms is adiabatically loaded into the optical lattice and the superfluid-to-Mott-insulator (SF-MI) transition is driven under conditions that maximize the number of doubly occupied lattice sites. Thus, a state is created in the central region of the optical lattice, with each lattice site filled with precisely two atoms \cite{jaksch2002}. The atom pair can be converted into a molecule in a well-defined rovibrational quantum state by magnetoassociation across a Feshbach resonance to create a Feshbach molecule \cite{herbig2003,regal2003}.  The coherent conversion of atom pairs into molecules can be considered as the ultimate control of a chemical reaction. Subsequently, the molecules are transferred to the desired internal state, most notably the lowest vibrational and rotational level of the electronic ground state, by coherent optical two-photon transitions. Two STIRAP (STImulated Raman Adiabatic Passage) steps \cite{bergmann1998} involving four laser transitions are used to efficiently transfer the molecules into the lowest rovibrational level of the ground state.

We theoretically investigate the interaction of cesium dimers with an external laser field with the aim to conveniently choose the wavelength of the optical lattice so that the initial and final molecular levels involved in the STIRAP sequence above are equally well trapped. At a "magic" wavelength \cite{katori1999}, the light shift for the two states of interest caused by the trapping light is equal, an important concept in precision metrology \cite{ye2008}. This enables the control over the motional wave function of the rovibrational ground state which is matched to the one of the initial state near threshold, thus avoiding a projection of the initial wave function onto higher motional states of the lattice during coherent manipulation of the internal state.  A similar strategy has recently be proposed with the addition of an external electric field, revealing "magic angles" between this field and the lattice field \cite{kotochigova2010}. Using accurate potential energy curves and transition dipole moments from accurate quantum chemistry computations \cite{aymar2011}, we calculate the dynamic dipole polarizability of cesium dimers via a summation over a large number of excited electronic states. We identified ranges of magic wavelengths where the ac Stark shift for the dimer in its final state, i.e the lowest rovibrational level of the ground state, is the same as for its initial state, i.e. a pair of non-interacting atoms.

This article is organized as follows: Section \ref{sec:method} recalls the basic definitions of the dynamic polarizability of a diatomic molecule. In Section \ref{sec:method} the relevant electronic properties of the Cs$_2$ molecule and the calculation of its dynamical polarizability is presented. Our results are presented in Section \ref{sec:alpha_cs2X} for the lowest rovibrational level of the $X^1\Sigma_g^+$ electronic ground state of Cs$_2$ and magic wavelengths conditions are investigated (Section \ref{sec:magic}). We generalize these calculations to the the lowest rovibrational level of the $a^3\Sigma_u^+$ lowest triplet state of Cs$_2$ (\ref{sec:alpha_cs2a}), before providing concluding remarks in Section \ref{sec:conclusion}. Atomic units (a.u.) will be used for distances (1~a.u.=0.0529177~nm) throughout the paper.

\section{Trapping with light and dynamic dipole polarizability}
\label{sec:method}

Following for instance Ref. \cite{grimm2000}, when an isolated atom or molecule is subject to an oscillating electric field  $\textbf{E}=\mathbf{\hat{e}}E_0\exp(-i\omega t)+\textrm{c.c.}$ with unit polarization vector $\mathbf{\hat{e}}$, frequency $\omega$, and intensity $I=2\epsilon_0c|E_0|^2$ ($\textrm{c.c.}$ holds for the complex conjugate), a dipole moment $\mathbf{p}=\mathbf{\hat{e}}p_0\exp(-i\omega t)+\textrm{c.c.}$  is induced oscillating at the same frequency. The response of the system to the field is characterized by the complex dipole polarizability $\alpha(\omega)$ according to $p_0=\alpha(\omega) E_0$. The related interaction potential is fixed by the real part of the polarizability $\Re (\alpha(\omega))$
\begin{equation}
U_{dip}=-\frac{1}{2} \langle \mathbf{p \cdot E} \rangle = - \frac{1}{2\epsilon_0c}\Re (\alpha(\omega))I
\label{eq:udip}
\end{equation}
where the angled brackets express the time average of the fast oscillating terms of $\textbf{E}$ and $\textbf{p}$. The imaginary part of the polarizability $\Im (\alpha(\omega))$ describes the absorption of the system through the power $P_{abs}$ absorbed by the oscillator
\begin{equation}
P_{abs}=\langle \mathbf{\dot{p} \cdot E} \rangle = 2\omega \Im (p_0E_0)= \frac{\omega}{\epsilon_0c}\Im (\alpha(\omega))I
\label{eq:pabs}
\end{equation}
These two quantities characterize the main properties of a dipole trap. The generic expression of the polarizability for a diatomic molecule in a state $|i>$ is
\begin{equation}
\alpha_i(\omega)=2\displaystyle {\sum_{f}}\frac{\omega_{if}-i\frac{\gamma_{f}}{2}}{(\omega_{if}-i\frac{\gamma_{f}}{2})^{2}-\omega^{2}}\left | \left \langle f \right | d(R)\hat{R}.\hat{\epsilon} \left | i \right \rangle \right |^{2}.
\label{eq:alpha}
\end{equation}
where the summation is covering all the accessible states $|f>$ with natural width $\gamma_f$ of the molecule through dipolar transitions with frequency $\omega_{if}$ characterized by the transition dipole moment $d(R) \hat{R}$, where $R$ is the internuclear distance and $\hat{R}$ the corresponding unit vector. This somewhat symbolic notation will be explained below, but we mention that the angled brackets refer to the spatial integration over all internal coordinates of the system. As is well known, an important feature of the optical dipole trap is the sign of the real part of the dynamic polarizability. In the case of a two-level system the situation is simple: when the laser frequency of the dipole trap is red detuned compared to the transition between the two states, the dynamic polarizability is positive and the dipole potential is attractive; in contrast, when the detuning $\Delta$ is to the blue, the dipole potential is repulsive. In the case of a multilevel system the situation is more complicated as resonances can compensate each other in the sum of Eq.(\ref{eq:alpha}), so that it is not obvious to predict when $\Re(\alpha)$ will be appropriate for trapping. We will see an illustration of this issue in the next sections.

The squared matrix element of Eq.(\ref{eq:alpha}) can be expressed in a more explicit way for a diatomic molecule relevant for the present study. Assuming that the $|i>$ ($|f>$) states are labeled with the symbols $\Xi_i$ ($\Xi_f$) for the electronic molecular state, $J_i$ ($J_f$), $M_i$ ($M_f$), and $\Lambda_i$ ($\Lambda_f$) for the total angular momentum and its projection on the $Z$ axis of the laboratory frame and on the molecular axis $z$, and $v_i$ ($v_f$) for the vibrational level, one finds for a linearly polarized field (along $Z$)
\begin{eqnarray}
\left | \left \langle f\right | d(R)\hat{R}.\hat{\epsilon} \left | i \right \rangle \right |^{2} = \nonumber \\
\left| \left \langle v_i\left |\displaystyle {\sum_{p=0,\pm1}} d_p^{if}(R) \left( \begin{array}{ccc} J_f & 1 & J_i \\
-M_f & 0 & M_i\end{array} \right) \left( \begin{array}{ccc} J_f & 1 & J_i \\
-\Lambda_f & p & \Lambda_i  \end{array} \right) \right | v_f \right \rangle \right|^2
\label{eq:tdm}
\end{eqnarray}
where $d_p^{if}(R)=\langle \Xi_i,\Lambda_i \left | \mu_p \right | \Xi_f,\Lambda_f \rangle$ is the matrix element of the component $\mu_p$ in the molecular frame $(xyz)$ of the electronic transition dipole moment at $R$, with $\mu_0=\mu_z$ and $\mu_{\pm 1}=\frac{1}{\sqrt{2}}(\mu_x\pm i\mu_y)$. The angled brackets in Eq.(\ref{eq:tdm}) refer to the integration on $R$, assuming that the vibrational wave functions are independent of the rotational level of the molecule, which is relevant for the low $J$ values investigated here. The non-vanishing matrix elements should fulfill the usual selection rules $\Delta \Lambda = 0,\pm1$ and $\Delta J = 0,\pm1$.

For the $X^1\Sigma_g^+$ ground state of an alkali-metal dimer ($\Lambda_i=0$) in a rovibrational level $J,M$, Eq.(\ref{eq:tdm}) reduces to
\begin{eqnarray}
\left | \left \langle f\right | d(R)\hat{R}.\hat{\epsilon} \left | i \right \rangle \right |^{2} = \, \\
\frac{2J^2+2J-1-2M^2}{(2J+3)(2J-1)} \left |\langle v_i \left | d_z^{if}(R) \right | v_f \rangle \right |^2 \nonumber
\label{eq:sigsig}
\end{eqnarray}
for $\Sigma-\Sigma$ transitions, and 
\begin{eqnarray}
\left | \left \langle f\right | d(R)\hat{R}.\hat{\epsilon} \left | i \right \rangle \right |^{2} =  \, \\
\frac{2J^2+2J-2+2M^2}{(2J+3)(2J-1)} \left |\langle v_i \left | d_x^{if}(R) \right | v_f \rangle \right |^2. \nonumber
\label{eq:sigpi}
\end{eqnarray}
for for $\Sigma-\Pi$ transitions. Therefore, the dynamic polarizability $\alpha_i(\omega)$ can be rewritten in the compact form
\begin{eqnarray}
\alpha_i(\omega) =\frac{2J^2+2J-1-2M^2}{(2J+3)(2J-1)}\alpha_{i \parallel}(\omega)+ \\
\frac{2J^2+2J-2+2M^2}{(2J+3)(2J-1)} \alpha_{i \perp}(\omega) \nonumber
\label{eq:alpha_compact}
\end{eqnarray}
where $\alpha_{i \parallel}(\omega)$ and $\alpha_{i \perp}(\omega)$ are the polarizabilities along the molecular axis (related to $\Sigma-\Sigma$ transitions) and perpendicular to the molecular axis (related to $\Sigma-\Pi$ transitions), respectively. If $J=0$ one finds the isotropic situation $\alpha_i=(\alpha_{i\parallel}(\omega)+2 \alpha_{i \perp}(\omega))/3$, which is also valid for any value of $J$ if the $M$ substates are statistically populated.

\section{Dynamic polarizabilities of ground state Cs$_2$ molecules}
\label{sec:alpha_cs2X}

\begin{figure}
\includegraphics[width=0.55\textwidth]{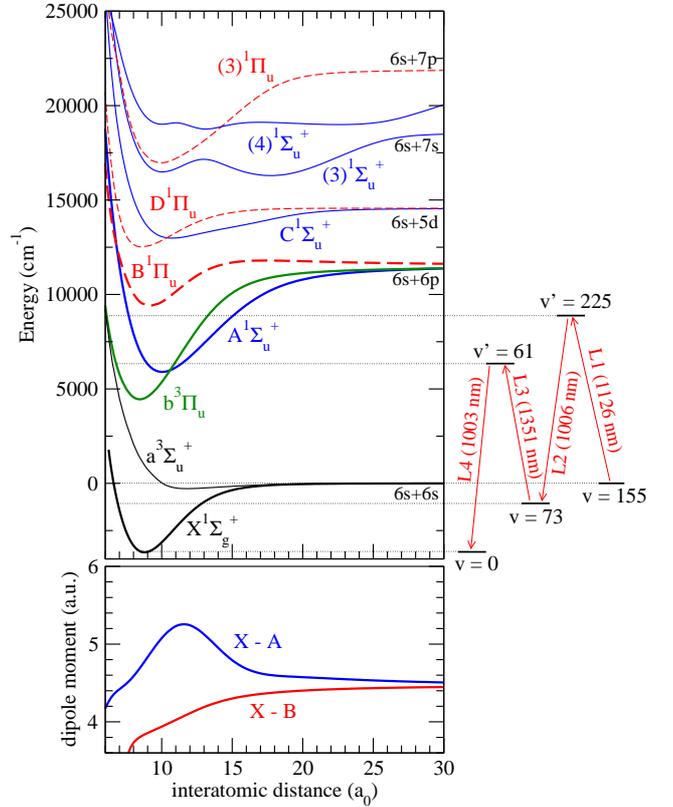}
\caption{\label{fig:pot} Ground and excited state potential energy curves (upper panel) and main transition dipole moments (lower panel) of Cs$_2$ as functions of the interatomic distance that are used for our calculations (see text). $X-A$ and $X-B$ refer to the $X^1\Sigma_g^+$-$A^1\Sigma_u^+$ and $X^1\Sigma_g^+$-$B^1\Pi_u$ transitions, respectively. The four-step STIRAP sequence used in the Innsbruck experiment is recalled for clarity.}
\end{figure}

As already stated, the calculation of $\alpha(\omega)$ following Eq.(\ref{eq:alpha}) requires the inclusion of all the molecular states $|f\rangle$ accessible via dipolar transitions from the initial state $|i\rangle$. For a Cs$_2$ molecule in a level ($v,J$) of its electronic ground state $X^1\Sigma_g^+$ (hereafter referred to as the $X$ state), all the rovibrational levels (including the continuum) of all electronic states of $^1\Sigma_u^+$ and $^1\Pi_u$ symmetries are needed, as well as the related transition dipole moments. An overview of the relevant molecular data is provided in Fig.\ref{fig:pot}, together with the levels involved in the four-step STIRAP process used to transfer the initial Feshbach molecules down to the ($v=0, J=0$) ground state level. The optimal efficiency of the motional control is achieved when both the Feshbach molecules and the ground state molecules are trapped in the 3D optical lattice, \textit{i.e.} their dynamic polarizability should be identical at the lattice wavelength.

\begin{figure}
\includegraphics[width=0.45\textwidth]{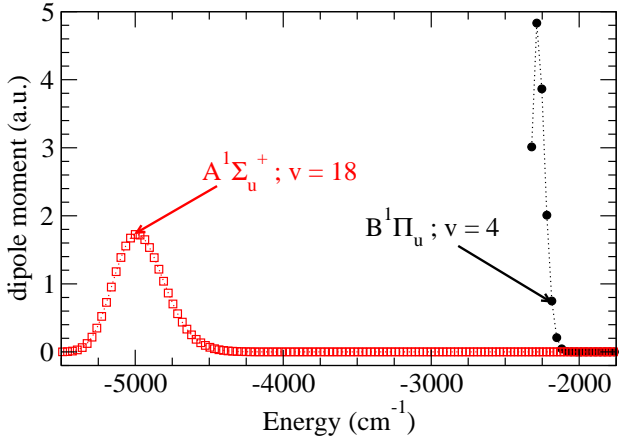}
\caption{\label{fig:overlap} Transition dipole moment from the $v=0$, $X^1\Sigma_g^+$ level to the levels of the first excited states $A^1\Sigma_u^+$ and $B^1\Pi_u$. The energy origin is taken at the $(6s+6p)$ dissociation limit.}
\end{figure}

\begin{figure}
\includegraphics[width=0.45\textwidth]{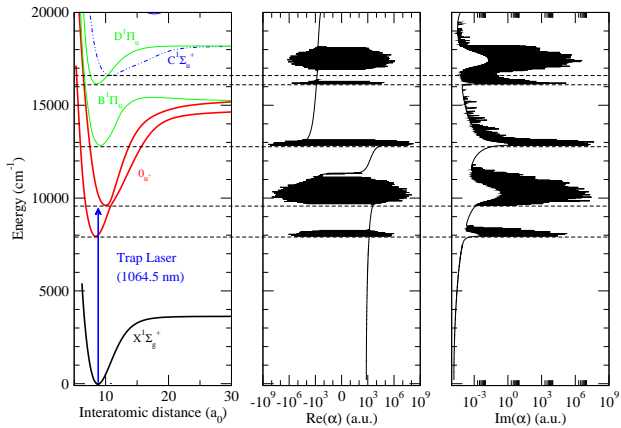}
\caption{\label{fig:dynpol_X} The ground and lowest excited potentials of Cs$_2$ (left panel), the real part (middle panel) and the imaginary part (right panel) of the dynamic polarizability of the $v=0$, $X^1\Sigma_g^+$ level as a function of laser energy. The polarizability is given on a logarithmic axis for both the positive and the negative part. The zero energy of the left panel is the energy of the $v=0$, $X^1\Sigma_g^+$ level. The horizontal dashed lines show the bottom position of each excited state to put into evidence the correlation between the resonances in the polarizability and the excited states.}
\end{figure}

Most of the potential energy curves (PEC) and all transition dipole moment (TDM) functions used here have been obtained in our group following the quantum chemistry (QC) approach described in Ref.\cite{aymar2005} and will be the subject of a separate publication. After convergence tests, the sum in Eq.(\ref{eq:alpha}) has been truncated to the three lowest $^1\Sigma_u^+$ states (correlated to the $6s+6p$, $6s+5d$, $6s+7s$ Cs$_2$ dissociation limits) and to the three lowest $^1\Pi_u$ states (correlated to the $6s+6p$, $6s+5d$, $6s+7p$ Cs$_2$ dissociation limits). We have also taken in account the available experimental information, for the $X$, $A^1\Sigma_u^+(6s+6p)$, and $B^1\Pi_u(6s+6p)$ states (hereafter referred to as the $A$ and $B$ state, respectively). We have used the Rydberg-Klein-Rees (RKR) PEC of Ref.\cite{amiot2002} for the ground state, and the RKR-PEC of Ref. \cite{diemer1991} between $R = 4$~a.u. and $R = 6$~a.u. matched to our QC calculations for the $B$ state. The $A$ state is coupled through spin-orbit (SO) interaction with the $b^3\Pi_u(6s+6p)$ (hereafter referred to as the $b$ state, see Fig.\ref{fig:pot}), giving rise to a pair of states of so-called $0_u^+$ symmetry correlated to the $(6s_{1/2}+6p_{1/2})$ and $(6s_{1/2}+6p_{3/2})$ limits, and exhibiting an avoided crossing in place of the crossing between $A$ and $b$. We have used the corresponding QC-PEC which we adjusted to reproduce (i) the energy position of the bottom of the experimental PECs as determined in Ref.\cite{verges1987} (for $A$) and in Ref.\cite{xie2008} (for $b$), (ii) the energy of the highly-perturbed levels observed in the Innsbruck experiments \cite{danzl2008,mark2009,danzl2009}. The molecular ($R$-dependent) SO coupling function is taken from the \textit{ab initio} determination of Ref.\cite{spies1989}. Note that a detailed spectroscopic analysis of the coupled $A$ and $b$ states has very recently become available \cite{bai2011}. All PEC have been smoothly matched to the long-range curves of Ref.\cite{marinescu1995} for completeness, but this had no influence on the final results. Finally, the vibrational energies and wave functions are computed with Mapped Fourier Grid Representation (MFGR) method \cite{kokoouline1999}. In Fig.\ref{fig:overlap}, the actual  TDMs obtained after integration on $R$ for the $X \rightarrow A$ and $X \rightarrow B$ vanish for high-lying vibrational levels, suggesting that the dissociation continua can be omitted in the sum of Eq.(\ref{eq:alpha}). Moreover, the TDMs towards levels of the other excited states are several orders of magnitude smaller than these ones.
The real part and imaginary part of the dynamic polarizability of a Cs$_2$ molecule in the $v=0$ level of its ground state \footnote{Note that $\Re(\alpha)$ includes the atomic core contribution as described in Section \ref{sec:magic}} are displayed in Fig.\ref{fig:dynpol_X} for a range of laser energies $E=\hbar \omega$ between 0 and 20000~cm$^{-1}$. They are evaluated with an energy step of 0.0125~cm$^{-1}$ (or 375~MHz), and assuming for simplicity a typical lifetime of 10~ns for all the excited states i.e $\gamma_f=15$~MHz. This assumption only influences the intensities of the resonance peaks visible in the polarizability, but not the magnitude of $\alpha(\omega)$ outside the resonant regions. The imaginary part is proportional to the line width of the excited states and thus to the absorption efficiency, which is found negligible outside the resonant regions, as expected. We see that the spin-orbit mixing of the $A$ and $b$ states induces resonances for excitation energies as low as 8000~cm$^{-1}$, \textit{i.e.} at the bottom of the lower $0_u^+$ state, so that light with photon energy smaller than this will safely lead to trapping. However, the Innsbruck experiment is performed with a 1064.5~nm laser (or 9394~cm$^{-1}$), which is to the blue of the transitions towards the levels of the lower $0_u^+$ state, and to the red of the levels of the upper $0_u^+$ state. Nevertheless, the calculation shows that there is indeed an optical window where the contributions of individual resonances all cancel out, so that the trapping of the molecules can be successful, in agreement with the observations. It is striking that other favorable windows (with $\Re(\alpha)>0$) exist at larger energies between resonant zones mainly occurring at the bottom of the wells. The region between 11200~cm$^{-1}$ and 12800~cm$^{-1}$ shows an abrupt change of sign of $\Re(\alpha)$  around 11200~cm$^{-1}$, which could be checked experimentally.

\section{Magic wavelength for optical trapping of ground state Cs$_2$ molecules}
\label{sec:magic}

The dynamic polarizability of a cesium atom can be written as \cite{derevianko2010}
\begin{equation}
\alpha_{Cs}(\omega)=\alpha_v(\omega)+\alpha_c(\omega)+\alpha_{cv}(\omega).
\label{eq:alpha_cs}
\end{equation}
The dominant contribution $\alpha_v$ comes from the valence electron and is calculated with Eq.\ref{eq:alpha}, where the energy of the atomic transitions and the transition dipole moments are taken from Ref.\cite{iskrenova2008}. The small contribution $\alpha_c$ involving core-excited state is chosen somewhat empirically as the difference at $\omega=0$ between our value for $\alpha_v$, and the one of Ref.\cite{derevianko2010} which indeed contains $\alpha_c$, yielding $\alpha_c=15.4$~a.u.. This value has been used for all frequencies far from core resonances, which is the case for an oscillating electric field corresponding to a trapping laser wavelength of 1064.5~nm, as used in the Innsbruck experiments. For this wavelength the valence polarizability is found equal to 1145.6~a.u. and the total atomic polarizability is then $\alpha_{Cs}=$1161~a.u.. Following Ref.\cite{derevianko2010}, $\alpha_{cv}$ is neglected compared to the other contributions.

The dynamic polarizability of a Feshbach molecule, \textit{i.e.} a molecule in a weakly-bound vibrational level of the ground state $X^1\Sigma_g^+$ near the dissociation limit $6s_{1/2}+6s_{1/2}$, is actually well approximated by two times the dynamic polarizability $\alpha_{Cs}$ of two ground state Cs atoms. If we describe the Feshbach molecule as a molecule in the highest lying vibrational level of the $X$ state, \textit{i.e}  ($v=155$, $J=0$), we find a value of 1.96$\alpha_{Cs}$ in good agreement with the measured one 2.02$\alpha_{Cs}$. We confirm that the dynamic polarizability of a Feshbach molecule is indeed very close to the one of an atom pair.

In the static case ($\omega=0$), the polarizability of the Cs$_2$ molecule in the ($v=0, J=0$) ground state level including the contribution of two Cs$^+$ cores ($2\alpha_c=30.8 $~a.u.) equals to 705~a.u., in good agreement with the one of Ref.\cite{deiglmayr2008} obtained by another approach \footnote{The value of Ref.\cite{deiglmayr2008} is given at $676.7$~a.u., and has to be augmented with twice the core polarizability, yielding 707.5~a.u.}.  This value could be expressed in units of Hz/[W/cm$^2$] which are more convenient for experimentalists since they allow to easily deduce the depth of the optical lattice for a given laser intensity: $1~a.u. = 4.6883572 \times 10^{-2}$~Hz/[W/cm$^2$].

\begin{figure}
\includegraphics[width=0.45\textwidth]{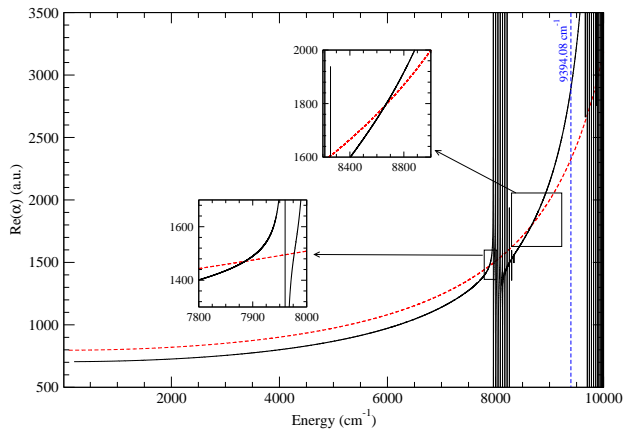}
\caption{\label{fig:magicX} Real part of the dynamic polarizability of the $v=0, J=0$, $X^1\Sigma_g^+$ level (black line) and twice the atomic polarizability (red dashed line) as a function of laser frequency. The vertical dashed line indicates the frequency of the optical lattice used in the experiments in Innsbruck (9394.08~cm$^{-1}$) corresponding to a wavelength of 1064.5~nm. }
\end{figure}

Figure \ref{fig:magicX} presents a zoom of $\Re(\alpha)$ around the region of interest for the Innsbruck experiment, plotted together with the real part of the  dynamic polarizability of a Cs atom pair simulating a Feshbach molecule. The figure reveals that both quantities are positive and have the same order of magnitude over a large energy range outside the resonant regions. They are found equal to 1787~a.u. (or 84~Hz/[W/cm$^2$]) at 8655~cm$^{-1}$, which is hence the so-called "magic" frequency for simultaneous optical trapping of both species. Note that there is another energy where both quantities are equal (around 7800~cm$^{-1}$) which is too close to the resonant region to be favorable.

We now compare the calculated and measured polarizabilities for molecules in $(v=0, J=0)$ at a trapping wavelength of 1064.5~nm. Our calculated value is found at 2.48~$\alpha_{Cs}$ while the measured one is found equal to 2.1~$\alpha_{Cs}$ \cite{danzl2010}. While in reasonable agreement, several features could explain this discrepancy. The calculated value relies on the precision of the PEC included in the sum of Eq.\ref{eq:alpha}, mainly of the states which have the predominant contribution, namely the $A/b$ coupled states and the $B$ states. The potential of the $B$ state is experimentally known over a tiny range of internuclear distances and thus can contribute to the inaccuracy of the computed value. Similarly, the PECs used for the $A/b$ system and the SO coupling function come from ab-initio computations, which are adjusted locally to reproduce some experimental data. This does not surely ensures that these data are correct for all internuclear distances. Finally the QC electronic transition dipole moments between the $X$ state and the  $A$ and $B$ states could also contribute to the discrepancy. Despite this discrepancy, we can conclude from Fig.\ref{fig:magicX} that the dynamic polarizabilities of the Feschbach molecule and of the ground state molecule are close enough over a sufficiently wide range of photon energies to allow favorable trapping conditions.

As an example of the possible anisotropy of the dynamic polarizability,  we have also calculated the polarizability of the $v=0, J=2$ level of the ground state at 1064.5~nm for different $M$ values, yielding $\alpha_{M=0}= 3907$~a.u., $\alpha_{M=\pm1}=3398$~a.u., $\alpha_{M=\pm2}=1870$~a.u.. The resulting average value is also found at 2.48$\alpha_{Cs}$. The energy of  each $(J, M)$ sublevel is shifted differently due to the quadratic Stark shift which is proportional to the polarizability acquired by the molecule in this sublevel. The sublevels $J, M$ are thus split by the electric field. With a field intensity of $10^3$~W/cm$^{2}$ typical of ongoing experiments, there is a total splitting of 95.5~kHz. This splitting due to the lattice light electric field has to be compared to other effects that can shift and/or split these levels such as the hyperfine structure and the Zeeman effect induced magnetic field. The anisotropic effects inducing the dependence of the polarizability versus frequency are not visible in the Innsbruck experiments where these effects are averaged since the optical lattice is a 3D one.

\begin{figure}
\includegraphics[width=0.45\textwidth]{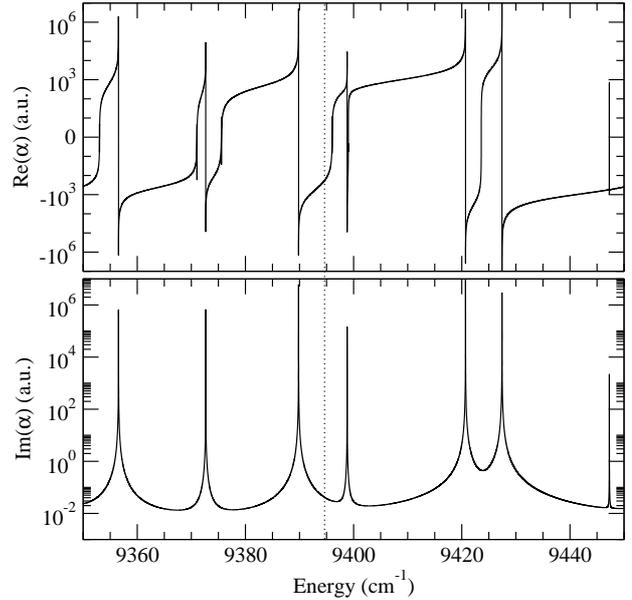}
\caption{\label{fig:alphaX73} Real and imaginary part of the dynamic polarizability of the $v=73$, $X^1\Sigma_g^+$ level  as a function of laser frequency. The vertical dotted line indicates for convenience the frequency of the optical lattice used in the experiments in Innsbruck (9394.08~cm$^{-1}$) corresponding to a wavelength of 1064.5~nm. }
\end{figure}

It is worthwhile to take a look at the dynamic polarizability of the $v=73$ level of the ground state. This level is the intermediate one in the 4-step STIRAP scheme. In the range of energy of the 1064.5~nm laser (Fig.\ref{fig:alphaX73}) lie many resonances so that the hope to temporally hold the molecules in the $v=73$ level could be quite tenuous. Due to the limited accuracy of the QC calculations, we cannot predict exactly which part of the polarizability function the lattice laser indeed reaches. However, a lifetime of 19~ms for the $v=73$ level in the dipole trap  has been measured \cite{danzl2009a}, which is much larger than the duration of the STIRAP sequence (shorter than 100~$\mu$s), but much shorter than the one for the Feshbach molecule or than the one for the $v=0$ level (several seconds). This is consistent with our calculations which suggest that even if the laser hits the resonant region, there are sufficient places in between the resonances to allow for a reasonable lifetime. Therefore it is clear that the magnitude of this polarizability is not important, as long as the trapping laser wavelength is not resonant with a level of the $A-b$ system.

\section{Dynamic polarizabilities of metastable triplet Cs$_2$ molecules}
\label{sec:alpha_cs2a}
\begin{figure}
\includegraphics[width=0.45\textwidth]{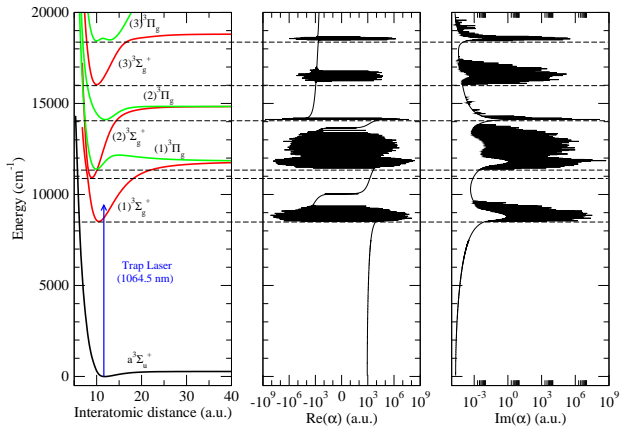}
\caption{\label{fig:dynpoltriplet} The potentials of the triplet states of Cs$_2$ involved in our calculations (left panel), the real part (middle panel) and the imaginary part (right panel) of the dynamic polarizability of the $v=0$, $a^3\Sigma_u^+$ level as a function of laser frequency. The polarizability is given on a logarithmic axis for both the positive and the negative part.}
\end{figure}
We have also performed the calculation of the dynamic polarizability for Cs$_2$ molecules in the ($v=0$, $J=0$) level of the lowest triplet state $a^3\Sigma_u^+$, which have been obtained for instance by photoassociation in the Orsay group \cite{viteau2008,viteau2009}. It is worthwhile to mention that such a population transfer applied to triplet Rb$_2$ trapped in an optical lattice has been already demonstrated \cite{lang2008a}.
The $a^3\Sigma_u^+$ state potential curve is taken from the recent analysis of Ref.\cite{xie2009}, while the potential curves of the triplet excited states and the $R$-dependent transition dipole moments are calculated in our group \cite{aymar2011}. After convergence checks, the calculation of the dynamic polarizability in the lowest triplet state involved six excited gerade triplet states allowed by the selection rules, the three lowest $^3\Sigma_g^+$ and the three lowest $^3\Pi_g$ states.

\begin{figure}
\includegraphics[width=0.45\textwidth]{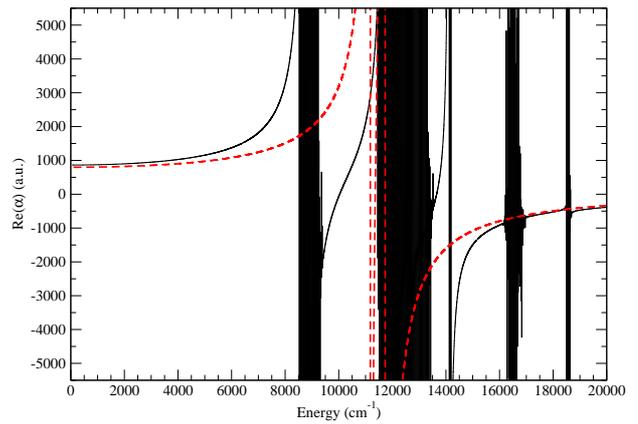}
\caption{\label{fig:magictriplet} Real part of the dynamic polarizability of the $v=0$, $a^3\Sigma_u^+$ level (black line) and twice the atomic polarizability (red dashed line) as a function of laser frequency. Far infrared laser should be adequate for trapping but there is no wavelength that exactly fulfills the magic wavelength condition outside the resonance zone.}
\end{figure}

We see in Fig. \ref{fig:dynpoltriplet} that the real part of the polarizability is smoothly varying over a broad range of frequencies red detuned to the first excited state ensuring attractive lattice potentials. In this frequency region, it is interesting to notice the smallness of the imaginary part of the polarizability ensuring small photon scattering rate. We see in Fig. \ref{fig:magictriplet} that the real part of the polarizability of the $v=0, J=0$ level of the triplet state is very close to the one of the atom pair simulating the relevant Feshbach molecule. We can see that the two curves never cross each other elsewhere than at two frequencies where at least one of those  polarizabilities shows resonance-like features. The difference between these two quantities in the long wavelength range outside the resonance region is even smaller than in the singlet case, so that we expect that the population transfer of the triplet molecules could be achieved in favorable trapping conditions, for instance around the readily available laser wavelength 1550~nm (or 6450~cm$^{-1}$).
\section{Conclusion}
\label{sec:conclusion}

In this work we have used accurate potentials curves for several singlet and triplet excited states and transition dipole moments to calculate dynamic polarizabilities acquired by Cs$_2$ molecules in the ground state and in the lowest triplet state when they interact with an oscillating laser field. These calculations have allowed us to find parameters for an optical lattice which optimizes the transfer of trapped Cs$_2$ molecules from an initial Feshbach state down to the lowest rovibrational level of the ground state, confirming the results obtained in the experiment of the Innsbruck team. In particular we found ranges of frequencies where the related dynamic polarizabilities are close enough together that the molecules are not excited into high motional modes of the lattice during the transfer, whatever their internal state is. We predict that there exists a magic wavelength for which the polarizabilities of both initial and final state of the transfer are equal. We also demonstrated that Cs$_2$ molecules created in a Feshbach level can be trapped simultaneously with molecules in the $v=0$ level of their lowest triplet state, yielding a good prospect for a STIRAP transfer in this case, just like it has been achieved with Rb$_2$ molecules \cite{lang2008a}. The generalization to the modeling of the trapping of RbCs molecules in order to design a similar transfer scheme down their lowest ground state level is under progress.

\section*{Acknowledgements}

We thank E. Haller for important contributions to the experimental work and R. Grimm for generous support. We acknowledge funding by the Austrian Science Fund (FWF) within project "Quantum Gases of Ground-State Molecules", project number P 21555-N20.


\end{document}